\documentclass[12pt,a4paper]{article}
\usepackage[utf8]{inputenc}
\usepackage[english]{babel}
\usepackage{amsmath}
\usepackage{amsfonts}
\usepackage{amssymb}
\usepackage{makeidx}
\usepackage{graphicx}
\usepackage{morefloats}
\usepackage[left=2cm,right=2cm,top=2cm,bottom=2cm]{geometry}
\author{Arindam Ghosh$^{1}$ and Sandip K. Chakrabarti$^{1,2}$}
\title{Anomalous Outbursts of H 1743-322}

\begin{document}
\maketitle
\begin{center}
{$^1$S. N. Bose National Centre for Basic Sciences, Salt Lake, Kolkata 700106, India.\\ 
$^2$Indian Centre for Space Physics, Chalantika 43, Garia Station Rd., Kolkata 700084, India.\\
{\it arindam.ghosh@bose.res.in; chakraba@bose.res.in}\\}
\end{center}
\abstract{Using black body and power-law photon counts of All Sky Monitor (ASM) in Rossi X-ray 
Timing Explorer (RXTE) satellite it has been established recently by us that 
there is a significant time lag between the infall timescales of two components
in the Two-Component Advective Flow (TCAF) paradigm, where a standard slow moving Keplerian disc 
is surrounded by a fast moving halo. The time lag is clearly due to the difference in viscosity 
in the flow components and the size of the Keplerian disc may be considered 
to be proportional to the arrival time lag. In this paper, using RXTE/ASM (1.5-12 keV) data, 
we examine eight successive outbursts of the low-mass X-ray binary H 1743-322 since 2003 from a new angle. 
A dynamic photon index, $\Theta$ indicates that the size of the Keplerian disc 
is biggest during the brightest outburst of 2003. The size diminishes thereafter during subsequent 
weaker outbursts. These results are corroborated when 
two energy fluxes corresponding to the two flows are cross-correlated 
with reference to $\Theta$. Moreover, $\Theta$ decides spectral 
transitions of any outburst. We show from the behaviour 
of $\Theta$ alone that the outburst of October 2008 was an anomalous 
outburst. In fact, each normal outburst was either preceded 
or followed by an otherwise premature outburst. 
This makes H 1743-322 an enigmatic source and a subject of further study.
}\\ 
%\smallskip


%%%%%%%%%%%%%%%%%%%%%%%%%%%%%%%%%%%%%%%%%%%%%%%%%%%%%%%%%

\section{Introduction}
Transient sources are low-mass X-ray binaries (LMXBs), which exhibit outbursts from time to time either regularly or sporadically. 
During a typical outburst, the soft (black body) photon flux peaks after a few days to a few weeks 
of the peaking of the hard (non-thermal power-law) photon flux (Ghosh \& Chakrabarti 2018).  
Moreover, the total radiated energy of outbursts of a single source can be very much different (Chakrabarti, Debnath, \& Nagarkoti 2018; Ghosh 2018; Ghosh \& Chakrabarti 2018; Zhou et al. 2013; Capitanio et al. 2009 and the references therein). 
It is believed that this variation is due to the variation of the accumulation radius, $X_{a}$.  At $X_{a}$, Keplerian disc matter 
is piled up due to a lack of viscosity in pre-outburst quiescent phase before being ushered in during outbursts
by viscosity higher than its critical value (Chakrabarti 1990, 1996). However, $X_{a}$ may be different for each outburst (Chakrabarti, Debnath, \& Nagarkoti 2018; 
Ghosh \& Chakrabarti 2018; Ghosh, Banerjee, \& Chakrabarti 2018). The accumulation radii represent the sizes of the Keplerian discs in multiple outbursts 
exhibited by transient LMXBs such as GX 339-4 or H 1743-322. It has been found earlier (Ghosh \& Chakrabarti 2018) that the disc size during 
the consecutive five outbursts of GX 339-4 varies from one to the other, even though both temporal and spectral properties are more or less similar. 
However, the successive outbursts of H 1743-322 are different in many ways from those in GX 339-4 (Ghosh 2018). For example, the outburst in 2008 is reported to be `a failed' (Capitanio et al. 2009) or `almost successful' (Zhou et al. 2013) since the source transits 
from hard state to hard-intermediate state before becoming harder again with decreased luminosity. 
H 1743-322 remained quiescent for about a quarter of a century until its detection in the outburst in March 2003 
and recently, Chakrabarti, Debnath, \& Nagarkoti (2018) show that the total energy liberated per outburst appears 
to be proportional to the time elapsed in quiescent state prior to the outburst, provided the outburst of 2004 is considered 
to be associated with that of the 2003.

In this paper, eight successive outbursts of H 1743-322 during 2003--2010 observed in RXTE era are analysed using the RXTE/ASM (1.5-12) keV data and within the framework of Two-Component Advective Flow (TCAF, Chakrabarti \& Titarchuk 1995). The two components in TCAF are responsible, in a non-linear 
way, for the two components of the observed spectra. A thin Keplerian disc having high angular 
momentum dissipates gravitational energy through viscosity and radiates multi-colour black body 
(soft) photons (Shakura \& Sunyaev 1973; hereafter SS73). A geometrically thick, optically thin, hot sub-Keplerian 
halo surrounds the Keplerian disc vertically. This component dissipates its thermal energy through recurrent 
inverse Comptonization close to the black hole, where it is puffed up due to the centrifugal barrier.
This region is known as the CENtrifugal pressure supported BOundary Layer, or CENBOL.
It produces high-energy (hard) photons as a power-law distribution in the spectrum. 
This emission property of a hot electron cloud or Compton cloud was demonstrated by Sunyaev \& Titarchuk (1980). 
Chakrabarti \& Titarchuk (1995) showed that the variation in X-ray flux from the Keplerian disc and the sub-Keplerian halo 
could be achieved by the changes in the mass accretion rates of the two components near the black hole. However,
the correspondence is not one to one, as the power-law photons are the results of repeated Compton scattering 
by the CENBOL and hence the CENBOL property is also important. However, the disc matter moves inward by 
viscosity. Unlike an SS73 disc, an advective sub-Keplerian flow with a low angular 
momentum does not require viscosity for accretion. As a consequence, the advective halo component 
arrives earlier than the viscous disc component. The difference in the times-of-arrival is defined as,
$$
\tau = t_{disc}-t_{halo},
$$
where, $t_{halo}$ and $t_{disc}$ are the infall timescales of the halo and the disc respectively. 
$\tau>0$ implies that the disc lags behind the halo. If $\tau=0$, it would mean that the disc arrives simultaneously with the halo.\\

According to transonic flow solution even if the accretion rate in the Keplerian disc is increased at the outer radius
the disc matter with high angular momentum would not be able to move inwards if viscosity 
is low enough to transport the angular momentum of the
enhanced Keplerian matter. The matter would therefore pile up at some radial distance $X_{a}$ 
until the flow becomes highly viscous, presumably due to convective viscosity, crossing the 
critical viscosity parameter (Chakrabarti, 1990, 1996). This will trigger a rapid inflow of 
disc matter and an outburst would ensue. If the viscosity is sufficient so that the spectra 
eventually becomes soft, we call it a normal outburst. In case 
the viscosity does not rise sufficiently, so as to evolve to either the hard-intermediate 
or the soft-intermediate state, or a premature hard state, the outburst will be anomalous 
or failed. Both the normal and anomalous outbursts should be readily distinguishable in the 
lightcurves of soft and hard radiation fluxes, or/and from their temporal (and spectral) 
characteristics. In general, the disc properties in H 1743-322 during all its outbursts can
be understood from a simple timing analysis discussed in $\S2$. Using the two energy 
fluxes (soft and hard) of X-ray radiation, and by the use of a suitably defined photon 
index, we show that the aforesaid travel time lag between the two components of accretion, 
and in turn, the disc size during the brightest/longest outburst (2003) in H 1743-322 
is much higher than those observed in its subsequent weak or faint outbursts. 
This is consistent with the preceding long time quiescence of the source 
(Chakrabarti, Debnath, and Nagarkoti 2018). The failure or an anomaly observed 
during the outburst of 2008 (Capitanio et al. 2009; Zhou et al. 2013) is more 
directly reflected from our analysis. Furthermore, each normal outburst is 
shown to be either preceded or followed by an otherwise premature or anomalous 
outburst. In $\S3$, we present the detailed results of our simple analysis. Finally, 
in $\S4$, we summarize major results.
  
\section{RXTE/ASM Data Analysis}

Despite the presence of negative counts and narrow energy binning, 
RXTE/ASM data is useful because of its availability for a long 
duration of time. During outbursts, these negative counts become very 
less in number. We use the {\sc ascii} version of public/archival 
RXTE/ASM lightcurve data during the outbursts of H 1743-322 
observed in the time span of about 7.5 years ($MJD~52700-MJD~55435$). 
RXTE/ASM operates over $1.5-12~keV$ energy range and it has three energy bands, viz. $A=(1.5-3 keV)$, $B=(3-5 keV)$, \& $C=(5-12 keV)$ respectively. If $a$, $b$, \& $c$ are respectively the number of photons in A, B, \& C bands with mean energies of $E_{A}=2.25~keV$, $E_{B}=4~keV$, \& $E_{C}=8.5~keV$ respectively, then the A-band represents low-energy flux $aE_{A}$ (or, soft flux) with absorbtion, B \& C bands together represent the total Comptonized, high-energy (3-12 keV) flux (or, hard flux) of $(bE_{B}+cE_{C})$.\\
%%%%%%%%%%%%%%%%%%%%%%%%%%%%%%%%%%%%%%%%%%%%%%%%%%%%%%%%%%%%%%%%%%
A dynamic photon index, $\Theta$, introduced and explained earlier in Ghosh \& Chakrabarti (2018), is given by, 
$$
\Theta = \tan^{-1} [\frac{(c-b)}{(E_{C}-E_{B})}].
\eqno{}
$$
Here, $\Theta$ is defined to be the slope of the hard region of the spectrum drawn in {\it linear scale}. 

By definition, $-1.57<\Theta<1.57$. In reality, $tan\Theta$ must be negative only, since $c$ is expected to be 
less than $b$ for equal energy binning. In soft states, $\Theta \rightarrow -1.57$ while in the hard states the spectrum is 
flatter, i.e., $\Theta \rightarrow 0$. However, in ASM, the size of the C band is largest, so the possibility of $c>b$ exists and this
often gives rise to $\Theta>0$ indicating a harder state. Physically, this spectral hardening happens due to a 
significant arrival time lag at the commencement of an outburst (Ghosh 2018; Ghosh \& Chakrabarti 2018). $\Theta$ is used for cross-correlation with individual fluxes, as the former represents relative change of one flux with respect to the other. So, the relative lag between the hard flux and the soft flux of either photons or energy would be defined as (Ghosh \& Chakrabarti 2018),
$$
\tau_{r}=t_{(\Theta,Soft)} - t_{(\Theta,Hard)}\rightarrow\tau.
$$          
$\tau_{r}>0$ implies that the disc radiation flux lags behind the halo radiation flux. 
Both evolve simultaneously if $\tau_{r}=0$. Typical behaviour of $\Theta$ during a 
normal outburst is schematically shown in Fig. 4 later. We wish to mention that while taking cross-correlations, the hard flux vs. 
soft flux correlation is not the proper quantity to discuss,
because the hard flux does not depend on the soft flux alone. 
It also depends on the properties of the Compton cloud such as its 
size (CENBOL) and its optical depth. Furthermore, in an outburst, 
hard flux will always be seen increasing with the soft flux. So the lag between hard and soft, in general, would be seen to be zero. What is important is the rate at 
which two accretion rates change, since they are guided by the 
viscosity in the halo and disc components in a complex way. 

Since RXTE/ASM did not record X-ray data continuously from any particular source, 
there are gaps in the data, in particular, due to the annual solar constraints. 
We smoothed out the unevenness of the data interval by using a {\sc fortran} 
code for interpolation. The {\sc fits} files obtained from these daily 
lightcurves are used to produce cross-correlations between $\Theta$ 
and two fluxes using {\sc crosscor} task of HEASOFT/XRONOS package in 
order to obtain the time lags estimated from $\Theta$. 

\section{Results}

We present the photon energy fluxes, $\Theta$-behaviour, 
their cross-correlations, and most importantly, the {\it hysteresis} 
property of the standard Keplerian disc for showing difference in 
temporal and spectral properties during various outbursts of 
H 1743-322. Figure 1 shows (a) total photon counts, (b) hard 
(dotted/online-blue) \& soft (continuous/online-red) photon 
fluxes, (c) average hard (dotted/online blue) \& soft (continuous/online-red) 
energy fluxes (intensities), and (d) the spectral slope $\Theta$. 
All the time series in Fig. 1 are drawn with weekly mean data for clarity. 
Eight successive outbursts are marked in (a) as OBn, where n=1, ..., 8. 
The outbursts are manifested by the strong enhancements in Fig. 1a and the strong dips 
in Fig. 1d. In addition, a flare-up soft state (which was not recognized as an outburst), 
marked as FUSS in Fig. 1a, 
is also considered in the same spirit, as the peak and total amount of 
soft flux are comparable to those in all outbursts. 
The spectral state becomes hard immediately after it, 
as is evident from the sharpest peak ($\sim$ MJD 53375) in $\Theta$ (Fig. 1d). 
Figure 2a shows bar charts or histograms, heights of which represent the 
total hard (dotted/online-blue) and soft (continuous/online-red) energy 
fluxes released per outburst. Equal width of the bars is not to 
the scale of time axis. Calendar years are also marked for 
referring to available reports. The dashed curve with triangles represents 
the ratio of soft energy flux to the hard one. This ratio is the highest 
($\sim 3.5$) in FUSS (2004), just following OB2 (2004), 
due to the lower value of total hard flux. However, in outbursts, 
this ratio never exceeds $0.3$. Histogram for FUSS is not shown 
here to avoid smudging. Triangles also indicate midway of 
the duration of outbursts. It is clearly seen from Fig. 2a that 
both soft and hard fluxes drop drastically in OB3 (2005) from their highest 
values in OB1 (2003). A long time ($>2$ years) of 
quiescence followed thereafter is instrumental behind the brighter OB4 (2007-8) 
and subsequent faint outbursts (OB5-OB8). Average hard energy flux (dotted/online-blue) 
and average soft energy flux (continuous/online-red curves) during the outbursts of 
Fig. 1 are zoomed in Fig. 3, where average daily data are used. 
Two peaks of the brightest outburst OB1 (2003) are marked as A \& B in (a). \\

Figure 4 is drawn schematically to show the typical behaviour of $\Theta$ during any normal outburst. 
This well-like appearance was earlier observed for ten normal outbursts in five other transient LMXBs, 
namely, GX 339-4, 4U 1543-47, XTE J1550-564, XTE J1650-500 \& GRO J1655-40 (Ghosh \& Chakrabarti 2018). 
Any normal outburst is characterized by four spectral states, viz. hard state (HS), 
hard-intermediate state (HIMS), soft-intermediate state (SIMS), and soft state (SS). 
Spectral transition must go through HS$\rightarrow$HIMS$\rightarrow$SIMS $\rightarrow$SS$\rightarrow$SIMS$\rightarrow$HIMS$\rightarrow$HS 
for a normal outburst. This is depicted in Fig. 4. $\Theta$ should attain its highest value at the beginning (HS) and at the end (HS) 
of a normal outburst. Approximately, $ \Theta \ge 0 $, $ \Theta_{\pm} \rightarrow -0.5 $, and $ \Theta \rightarrow -1 $ 
can be respectively assigned to HS, HIMS/SIMS, \& SS for a better understanding of our results. 
In our analysis, $\Theta_{maximum}$ always exceeds $0$ in HS, $\Theta_{minimum}$ does not go below $-1$ in SS except for OB1 (2003). 
The possibility of having more hard/soft state(s) within the well of Fig. 4 would indicate the multiplicity (twin, trine, etc.) 
of the outburst, or the number of mini outbursts. If any outburst shows significant deviations of $\Theta$ behaviour from that of Fig. 4, then we will call the outburst as `anomalous'. 

Figure 5 shows the actual behaviour of $\Theta$ with time during all outbursts, which are labelled as OB1--OB8 along the vertical axes. The dotted/online-red curves are drawn to represent average envelopes of 
$\Theta$ so that the aforesaid multiplicity becomes clearer. $\Theta$ attains its 
highest value a few days after the commencement of the normal outbursts OB1 
(2003), OB2 (2004), OB3 (2005), \& OB6 (2009) and rises again at their culmination, 
as seen in Figs. 5a, 5b, 5c, \& 5f respectively. On the other hand, its trend is the same 
as above at the onset of OB5 (2008) in Fig. 5e, but it declines thereafter. 
With reference to Fig. 4, a direct HS$\rightarrow$ HIMS spectral transition 
and a further hardened spectrum at the end of this {\it failed} 
outburst are indicated (Capitanio et al. 2009), whereas in view 
of Figs. 5c \& 5f it is {\it almost successful} (i.e. almost normal) outburst (Zhou et al. 2013). 
In view of both Fig. 3a \& Fig. 5a, OB1 (2003) is a twin outburst characterized by two soft states 
(though OB1-A could be a twin itself due to an interim temporary HS) and
OB1-B is normal and comparable to the following OB2 of Fig. 5b. However, 
features of $\Theta$, in comparison with Fig. 4, are deviated in (Fig. 5d) 
OB4 (2007-8), (Fig. 5g) OB7 (2009-10) \& (Fig. 5h) OB8 (2010). Erratically fluctuating 
$\Theta$ is probably due to multiply exhibited mini outbursts (indicated by multiple wells 
in dotted/online-red) with rapid softening/hardening spectra.
There appears to be same spectral states occuring several times in the same outburst.

It has been shown in Ghosh \& Chakrabarti (2018) that the commencement of an outburst in HS does not ensure an immediate spectral transition 
to a softer state. Rather, the spectrum becomes 
gradually harder. According to TCAF, as the halo matter rushes in, the hard photons become spectrally harder with more electrons 
to cool down. The time elapsed in attaining the hardest possible state from HS would indicate the lag time between disc and the halo components. These time lags (ranging from $13d$ to $1d$) are marked in Figs. 5a, 
5c and 5e-g. In Figs. 5b \& 5d, the time lag is approximately $0d$. 
However, this lag time is not generally observed from a direct flux-flux cross-correlation (CC) method, but readily evident from $\Theta$-flux correlations as illustrated in Fig. 6. CC($\Theta$, Hard Flux) and CC($\Theta$, Soft Flux) 
are represented respectively by dotted/online-blue and continuous/online-red curves. 
Dashed curves show CC(Hard Flux, Soft Flux). Two regions A \& B of 
Figs. 3a \& 5a are used separately in Fig. 6a. Non-zero time lags are marked in each 
box. A lag time of about a week, as indicated in Fig. 5f for OB6 (2009), 
is consistent with a lag of $5\pm3.5d$ obtained from $\Theta$-flux 
correlations (Fig. 6f). Similarly, a lag of about two weeks in 
Fig. 5a for OB1 (2003) is corroborated by $20\pm8d$ from Fig. 6a. 
So, we can rely upon $\Theta$-behaviour alone in understanding 
all the outbursts. This makes $\Theta$ a {\it spectro-temporal} 
index which dynamically traces any outburst throughout in terms 
of both spectral and temporal characteristics. Figs. 6e \& 6g 
are different from the rest of the Figures: 
hard and soft fluxes are anti-correlated with reference to 
$\Theta$. An exceptional lag of $11d$ in Fig. 6g for OB7 (2009-10) 
from direct CC(Hard Flux, Soft Flux) may be due to either the fact that
fresh incoming matter with the leftover of OB6 (2009) increased hard 
flux, or a jump in Comptonization efficiency following a sudden 
increase in accretion rate (see also Figs. 3g \& 5g). This further 
ensures that a fast increase in hard flux does not guarantee 
a spectral transition. Although the brightest outburst in Fig. 6a 
shows a time lag of $20\pm 8d$ for OB1-A (2003), but the lags 
are negligibly small in the subsequent weaker outbursts.

In Fig. 7, mean intensity in all the outbursts of Fig. 3 is plotted 
with $\Theta$ of Fig. 5. However, without any loss of generality, 
curves are drawn with running average data for clarity and 
comparison. Time lags obtained in Fig. 6 are marked along the vertical axis of 
each box. The multiplicity observed again in closed ({\it hysteresis}) loops is reasonable 
for the number of wells in $\Theta$ (Fig. 5). Closed loops in (a), (b), (c) \& (f) are respectively pertaining to normal 
outbursts OB1 (2003), OB2 (2004), OB3 (2005) \& OB6 (2009). These are consistent 
with the $\Theta$-behaviour in Figs. 5a-c \& 5f respectively. 
Loops in solid and dotted/online-red in Fig. 7a pertain to OB1-A \& OB1-B respectively. 
OB1 (2003) is a twin outburst, though OB1-A itself is complex showing an anomaly and is itself a twin due to 2-fold loop, arisen due to the presence of an interim temporary HS (Fig. 5a); OB1-B is comparable to OB2 of Fig. 6b and both are normal. For anomalous outbursts, OB4 (2007-8), OB7 (2009-10) \& OB8 (2010), the excursion 
path is erratic as seen respectively from Figs. 7d, 7g \& 
7h. The multiplicity is also apparent from the folded loops (see also Figs. 5d, 5g \& 5h). 
The closed loop in Fig. 7e indicates that OB5 (2008) could almost be a normal outburst which lacks a proper softening as $\Theta>-0.3$ (see Fig. 4 \& Fig. 5e). The biggest loop area in Fig. 7a (for OB1-A) is consistent with the corresponding highest 
time lag of about $20 \pm 8d$. This indicates the largest disc size 
in the brightest outburst OB1 (2003) in H 1743-322. 
Negligibly small lag in all subsequent outbursts is indicative of small size of the Keplerian disc.

Figure 8 shows (a) energy fluxes, (b) $\Theta$, (c) cross-correlation profile and (d) hysteresis diagram for FUSS (2004). Line types or colour stamps are similar to Figs. 3, 5 \& 6. The magnitude of soft flux in (a) is comparable with those in all outbursts. FUSS is also showing a striking excursion path in (d). The lumpy behaviour of $\Theta$ and hysteresis diagrams (Figs. 5d, 5g, 5h, 7d, 7g \& 7h) indicate that  OB4 (2007-8), OB7 (2009-10) \& OB8 (2010) are comparable and almost similar to FUSS (2004). These three, along with OB5 (2008), are possibly premature outbursts. Chakrabarti et al. (2018) have recently shown that OB2 (2004) could also partially release energy from disc matter piled up in the period
prior to OB1 (2003). Similarly, FUSS (2004) can also arise from the surplus disc matter of OB2 (2004).

It is expected that due to viscous time needed to transport matter inward, farther the piling 
radius $X_{a}$ is, greater should be the time lag. OB2 (2004) is a normal outburst 
for which $X_{a}$ was very close to the outer edge of the CENBOL because 
of $0d$ time lag. FUSS (2004) occurs immediately after OB2 (2004). OB3 (2005) is 
normal followed by anomalous OB4 (2007-8). Then appears `almost normal' OB5 (2008), which precedes the next normal OB6 (2009). The next one, OB7 (2009-10), is once again anomalous. This is followed by yet another anomalous OB8 (2010). All numerical estimates of our analysis are summarized in Table 1, which also briefly indicates the transient trend in H 1743-322. Therefore, by treating FUSS (2004) as a mini-outburst, every normal outburst following the brightest OB1 (2003) appears to be either followed or preceded by one or more otherwise premature outbursts with varying degrees of anomaly and transience.
 
\section{Summary}

In this paper, we have revisited the properties of several outbursts from a completely different perspective in order to have a unifying understanding of this unique system. After about two and a half decades, in 2003, an outburst in H 1743-322 occurred, which released enormous energy and had a very complex structure. Soon after in 2004, again another outburst occurred which Chakrabarti et al. (2018) considered as a leftover of the energy that should have been emitted in 2003. After this till 2010, outbursts, which we considered, were occurring at almost regular intervals and releasing a very little energy compared to that of 2003. In a `normal' outburst, the spectral state usually passes through the HS, HIMS, SIMS, and SS before returning back to the HS following the reverse route. Some of the outbursts did not even reach the SIMS or SS. In Mondal et al. (2017, and references therein) it was mentioned that when the viscosity parameter at the piling radius at a large distance crossed well above the critical value, the disc would become Keplerian till the inner edge and a soft state could form. If the viscosity did not rise much, the outburst would be anomalous. It has been shown that the viscosity parameter indeed has a specific limit to cross in order to reach a certain spectral state.

We could differentiate the outbursts in H 1743-322 by a completely new method, and in that process, obtain new insights if we stick to the TCAF paradigm. Here, the Keplerian disc of high viscosity is sandwiched by the sub-Keplerian flow having low viscosity and thus two types of time lags between the hard and the soft radiation fluxes are expected. First, a lag comes from the fact that as soon as an outburst is triggered at the accumulation radius, the sub-Keplerian component matter rushes towards the black hole and the spectrum hardens since more electrons are to be cooled down by the soft photons. The times taken to reach the hardest spectra would be proportional to the size of the disc. 
The second lag comes from the fact that the Keplerian matter reaches much after that through viscous time scale. 
In general, both the fluxes start to rise, and thus direct cross-correlation between the hard and soft fluxes would not give any information about these lags. However, the spectral slope $\Theta$ that we defined, continuously traces the way one component changes with respect to the other, and thus the cross-correlations of the fluxes with $\Theta$ would reveal the lags.

We have computed these cross-correlations for all the outbursts and we could easily 
identify that the initial lag for further hardening 
is different for different outbursts. This would indicate that the accumulation radius is
also different. Second, we have found that for 
a normal outburst, the mean intensity vs. $\Theta$ makes a simple closed loop.
On the contrary, for an anomalous outburst, this becomes complex with multiple/folded loops. Third, we found that the 2003 outburst appeared to consist of (at least) two separate outbursts occurring back to back. 
There was a `flare' type outburst, which we termed as FUSS, 
whose behaviour was different from the others in 
that the hardness ratio behaves in an opposite manner. Moreover, when seen together, there is a general tendency to have a normal outburst followed by one or two anomalous ones. 
Compared to this object, in GX 339-4, which also exhibits several 
outbursts, such anomalies were not observed (Ghosh \& Chakrabarti 2018; Ghosh 2018). 
However, there also, it was found that $X_{a}$ could go farther away or 
come closer in successive outbursts in GX 339-4. In H 1743-322, $X_{a}$ 
seems to be at the farthest distance prior to OB1 (2003) 
with the largest disc. Thereafter, $X_{a}$ moves  closer with 
smaller disc sizes. However, our study revealed that not all 
the matter piled up in the preceding quiescent state is always 
emptied out at the outburst. Multiple attempts could have been made to 
evacuate the matter (such as twice/thrice in 2003 and once in 2004), if the piled up matter 
is very high. We also find that the normal outbursts are separated by 
anomalous ones. We shall continue such studies in other systems having 
recurring outbursts to understand the physics behind outbursts in black hole binaries.\\  

%%%%%%%%%%%%%%%%%%%%%%%%%%%%%%%%%%%%%%%%%%%%%%%%%%%%%%%%%%%%%%%%%%%%
\section*{Acknowledgement}
The authors are indebted to NASA Archives for RXTE/ASM public data and facilities.
%%%%%%%%%%%%%%%%%%%%%%%%%%%%%%%%%%%%%%%%%%%%%%%%
%%%%%%%%%%%%%%%%%%%%%%%%%%%%%%%%%%%%%%%%%%%%%%%%%%%%%%%%%%%%%%%%%%%%%%%%%%

%%%%%%%%%%%%%%%%%%%%%%%%%%%%%%%%%%%%%%%%%%%%%%%%%%%%%%%%%%%%%%%%%%%%%%%%%%%%
%%%%%%%%%%%%%%%%%%%%%%%%%%%%%%%%%%%%%%%%%%%%%%%%%%%%%%%%%%%%%%%%%%%%%
%%%%%%%%%%%%%%%%%%%%%%%%%%%%%%%%%%%%%%%%%%%%%%%%%%%%%%%%%%%%%%%%%%%%%%%%%%%
\newpage
%\begin{landscape}
%%%%%%%%%%%%%%%%%%%%%%%%%%%%%%%%%%%%%%%%%%%%%%%%%%%%%%%%%%%%%%
\begin{table}
\begin{center}
\caption{Estimation of $\tau_{r}$ and Characteristics of Outbursts in H 1743-322}
\vspace*{4 mm}
%{\tiny
\begin{tabular}{c|c|c|c|c|c|c|c}
\hline\noalign{\smallskip}
%\hline
%&&&&\\
 Time Span & Calendar & Outburst & $\tau_{r}$ (d) & $\tau_{r}$ (d) & $\tau_{r}$ (d) & Remarks & Disk\\
(MJD) & Year & & ($\Theta$) & ($\Theta$-Flux) & (Flux-Flux) & & Size\\
%&&&&\\
\hline
& & (A) & 13 & $20 \pm 8 $ & $ 0 \pm 1$ & Anomalous & Big \\
 $52720-52950$ & 2003 & OB1 & & & & Brightest  & \\
 & & (B) & -- & $2 \pm 0.7$ & $ 0\pm 1$ & Normal & Small \\
 \hline
 $53181-53300$ & 2004 & OB2 & 0 & $0 \pm 1.1$ & $0\pm 1$ & Normal & Small \\
 $53350-53380$ & 2004 & FUSS & 4 & $4 \pm 0.7$ & $4\pm 2$ & Anomalous & Small \\
 $53588-53641$ & 2005 & OB3 & 2 & $2 \pm 1.4$ & $0\pm 1$ &  Normal & Small \\
 $54430-54490$ & 2007-8 & OB4 & 0 & $2\pm 7.8$ & $0 \pm 1$ & Anomalous & Small \\
%\hline
 $54735-54772$ & 2008 & OB5 & 3 & $0 \pm 6.4$ & $0 \pm 7$ & Anomalous & Small \\
 %\hline
 $54970-55020$ & 2009 & OB6 & 7 & $5 \pm 3.5$ & $2 \pm 1$  & Normal & Moderate \\
 %\hline
 $55165-55240$ & 2009-10 & OB7 & 2 & $0 \pm 0.7$ & $11 \pm 1$  & Anomalous & Small\\
$55415-55435$ & 2010 & OB8 & 1 & $1 \pm 3.2$ & $0 \pm 2$ & Anomalous & Small\\
%&&&&\\
\hline
\end{tabular}
%}
\end{center}
%\hspace*{4 mm}
%\renewcommand{\thefootnote}{\alph{footnote}}
%\let\thefootnote\footnote{$^*$}
\end{table} 
%\end{landscape}

%%%%%%%%%%%%%%%%%   TABLE ENDS HERE   %%%%%%%%%%%%%%%%%%

\begin{figure}
\begin{center}
\includegraphics[width=\columnwidth]{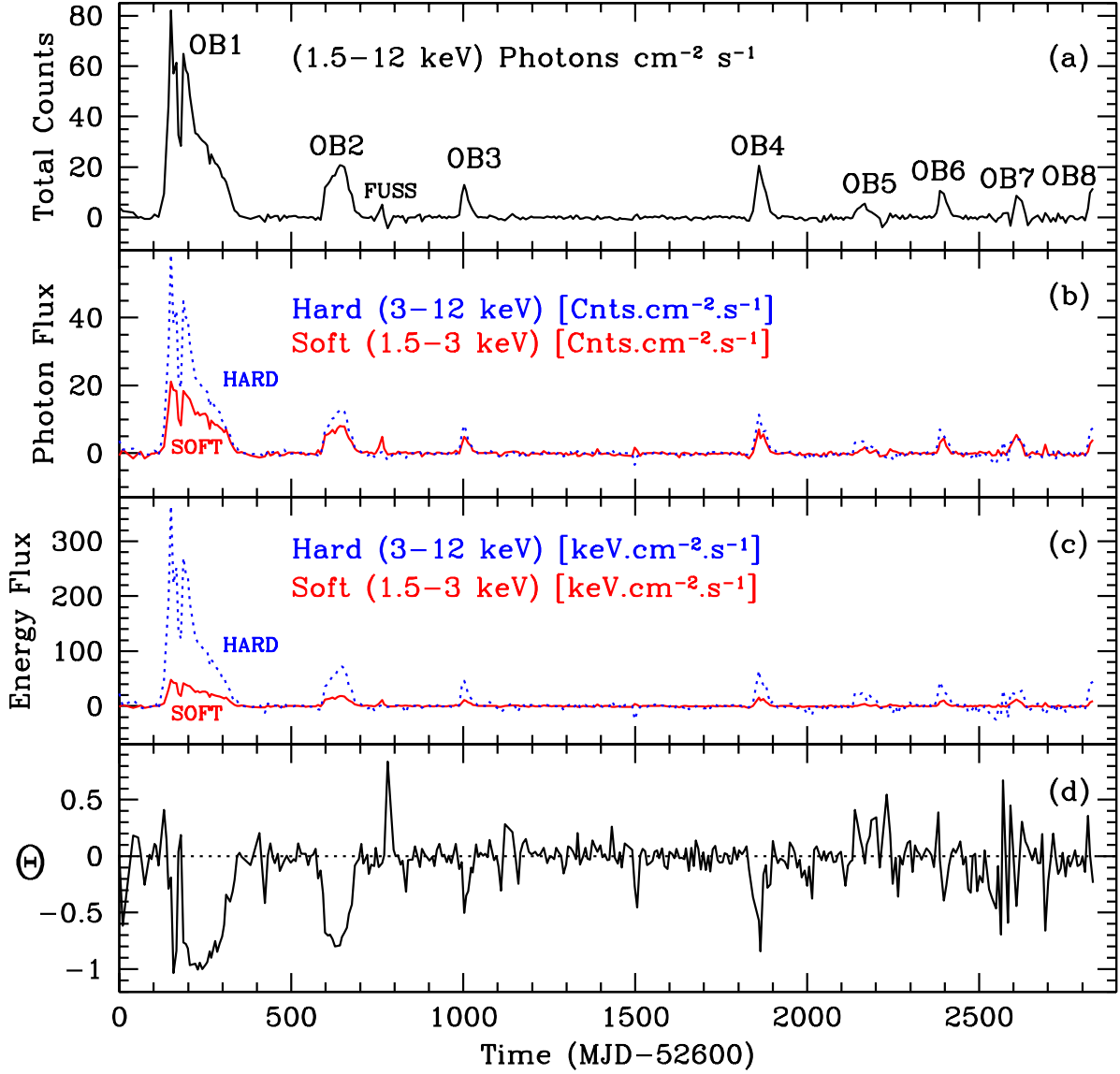}
\caption{(a) Total photon counts, (b) hard (dotted/blue) \& soft (continuous/red) photon fluxes, (c) average hard (dotted/online-blue) \& soft (continuous/online-red) energy fluxes (intensities), and (d) the slope $\Theta$. All time series are drawn with week-mean data for clarity. Eight successive outbursts are marked in (a) as OBn, where n=1, ..., 8. These outbursts are manifested by the strong dips in (d). In addition, a flare-up soft state, marked as FUSS in (a), is also considered in the same spirit, as its peak and total amount of soft flux are comparable to those in all OBs; the spectral state becomes hard immediately after it, as is evident from the sharpest peak ($\sim$ MJD 53375) in (d).}
\end{center}
\end{figure}
%%%%%%%%%%%%%%%%%%%%%%%%%%%%%%%%%%%%%%%%%%%%%%%%%%%%%%%%
%%%%%%%%%%%%%%%%%%%%%%%%%%%%%%%%%%%%%%%%%%%%%%%%%%%%%%%%%
\begin{figure}
\begin{center}
\includegraphics[width=\columnwidth]{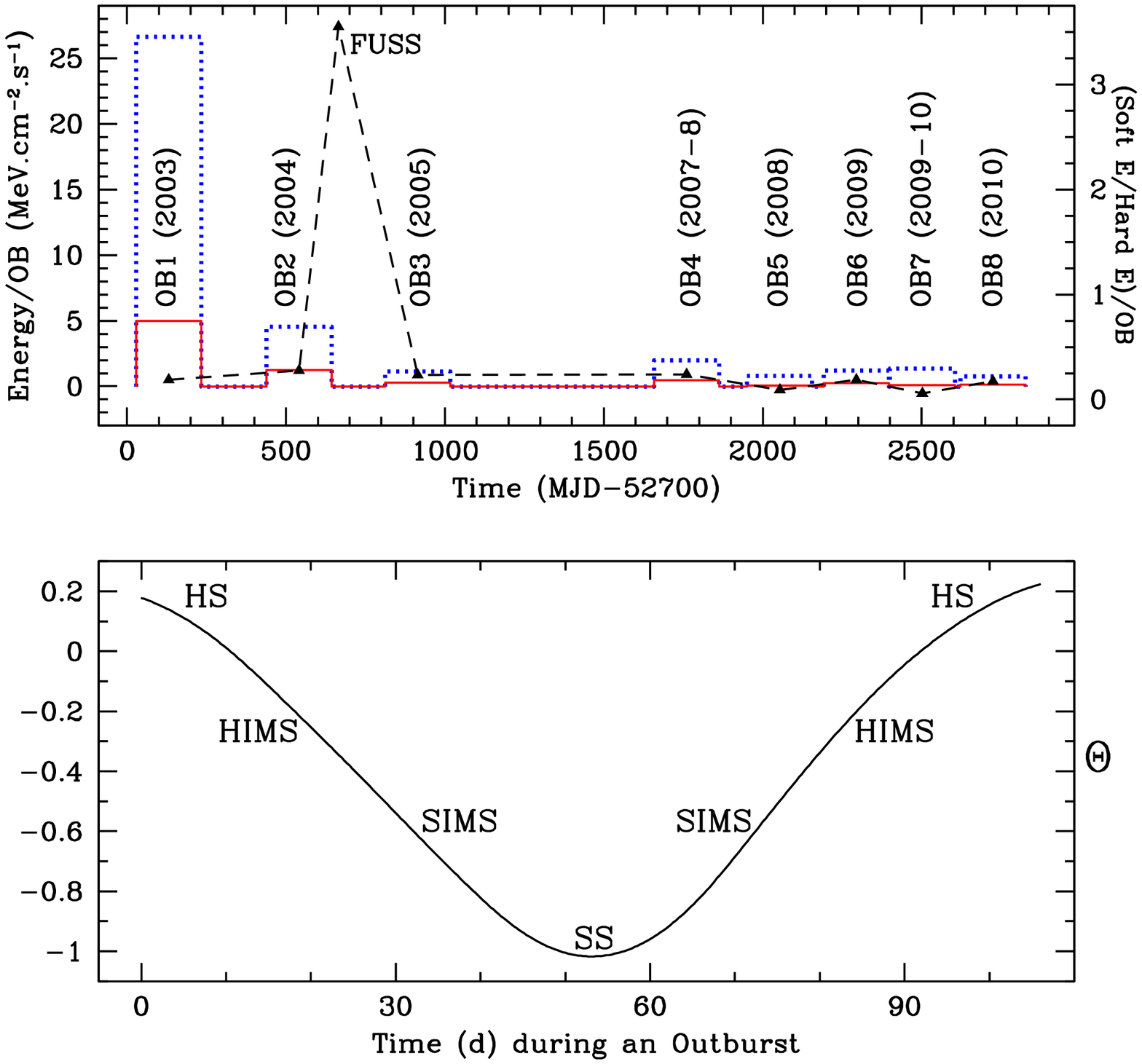}
\caption{Bar charts or histograms represent the total hard (dotted/online-blue) and soft (continuous/online-red) energy fluxes released per outburst by their heights. Equal width of the bars has nothing to do with the duration. Calendar years are also marked. The dashed curve with triangles represents the soft-to-hard ratio of these fluxes; this ratio is highest in FUSS (2004), the histogram of which is not shown here. Triangles also represent the central points during outbursts.}
\end{center}
\end{figure}
%%%%%%%%%%%%%%%%%%%%%%%%%%%%%%%%%%%%%%%%%%%%%%%%%%%%%%%%%

\begin{figure}
\begin{center}
\includegraphics[width=\columnwidth]{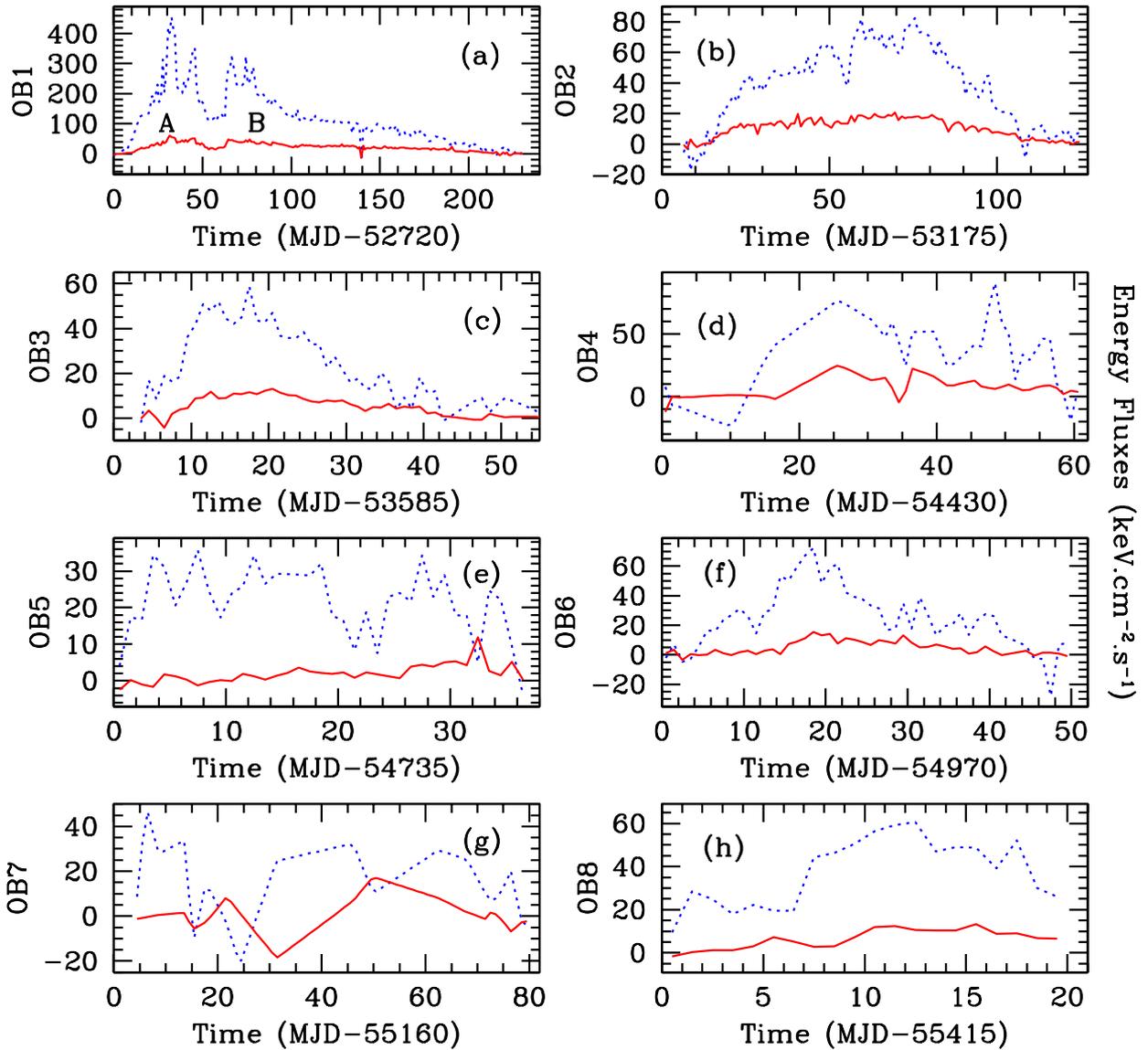}
\caption{Average hard energy flux (dotted/online-blue) and average soft energy flux (continuous/online-red) during the outbursts of Fig. 1 are zoomed. Average daily data are used. Two parts showing two peaks of the brightest outburst OB1 (2003) are marked as A \& B in (a).}
\end{center}
\end{figure}

%%%%%%%%%%%%%%%%%%%%%%%%%%%%%%%%%%%%%%%%%%%%%%%%%%%%%%%%%
\begin{figure}
\begin{center}
\includegraphics[width=\columnwidth]{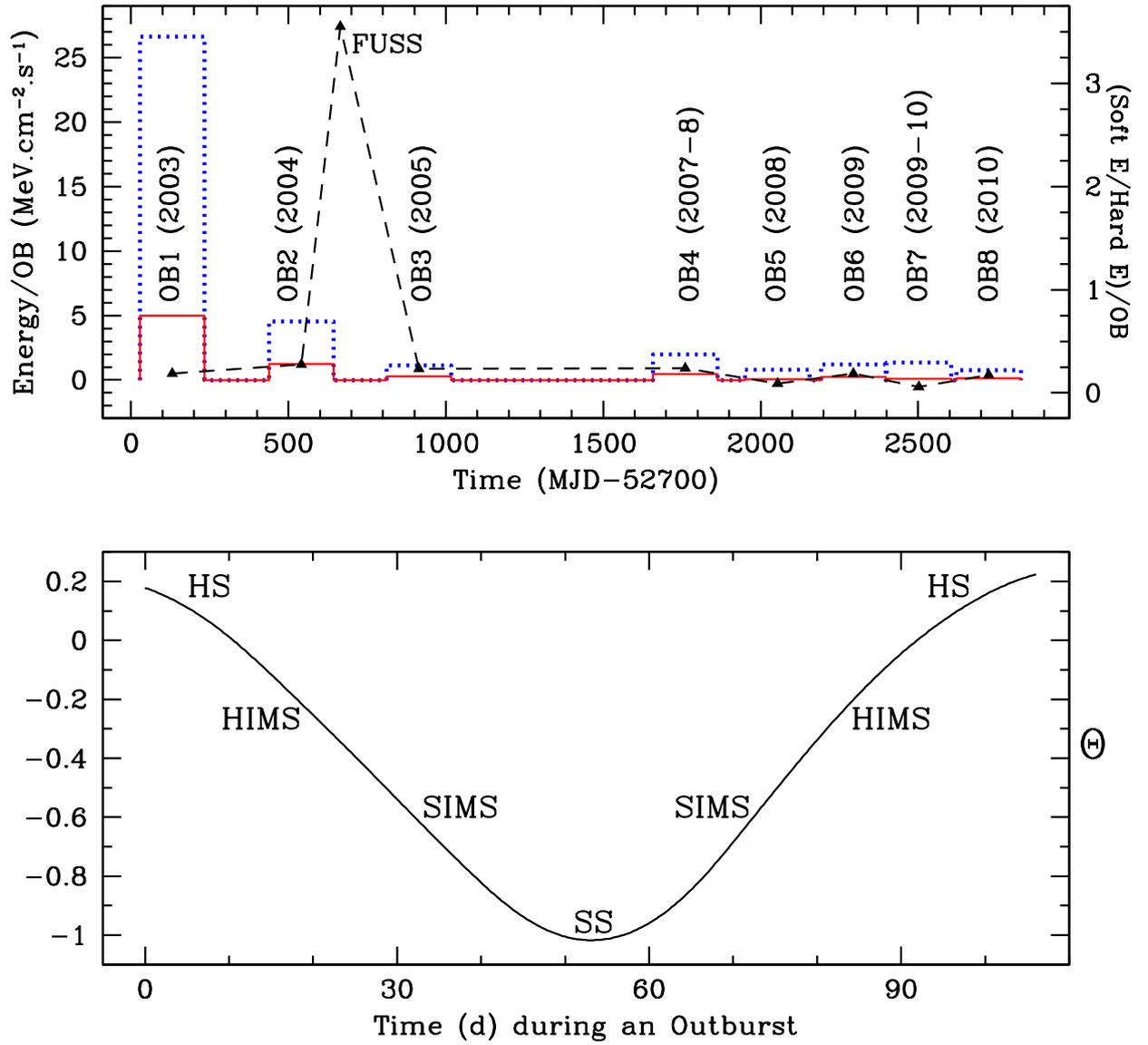}
\caption{Typical behaviour of $\Theta$ in any normal outburst is shown by the well-like schematic curve. Hard state (HS), hard-intermediate state (HIMS), soft-intermediate state (SIMS), and soft state (SS) are approximately marked. $\Theta$ should be highest at the beginning (HS) and at the end (HS) of a normal outburst. The typical ranges of $\Theta$, namely, $ \Theta \ge 0 $, $ \Theta_{\pm} \rightarrow -0.5 $, and $ \Theta \rightarrow -1 $ can be assumed for HS, HIMS/SIMS, and SS respectively.}
\end{center}
\end{figure}
%%%%%%%%%%%%%%%%%%%%%%%%%%%%%%%%%%%%%%%%%%%%%%%%%%%%%%%%%%%%%%%%%%%%
%%%%%%%%%%%%%%%%%%%%%%%%%%%%%%%%%%%%%%%%%%%%%%%%%%%%%%%%%%%%%%%%%%%%
\begin{figure}
\begin{center}
\includegraphics[width=\columnwidth]{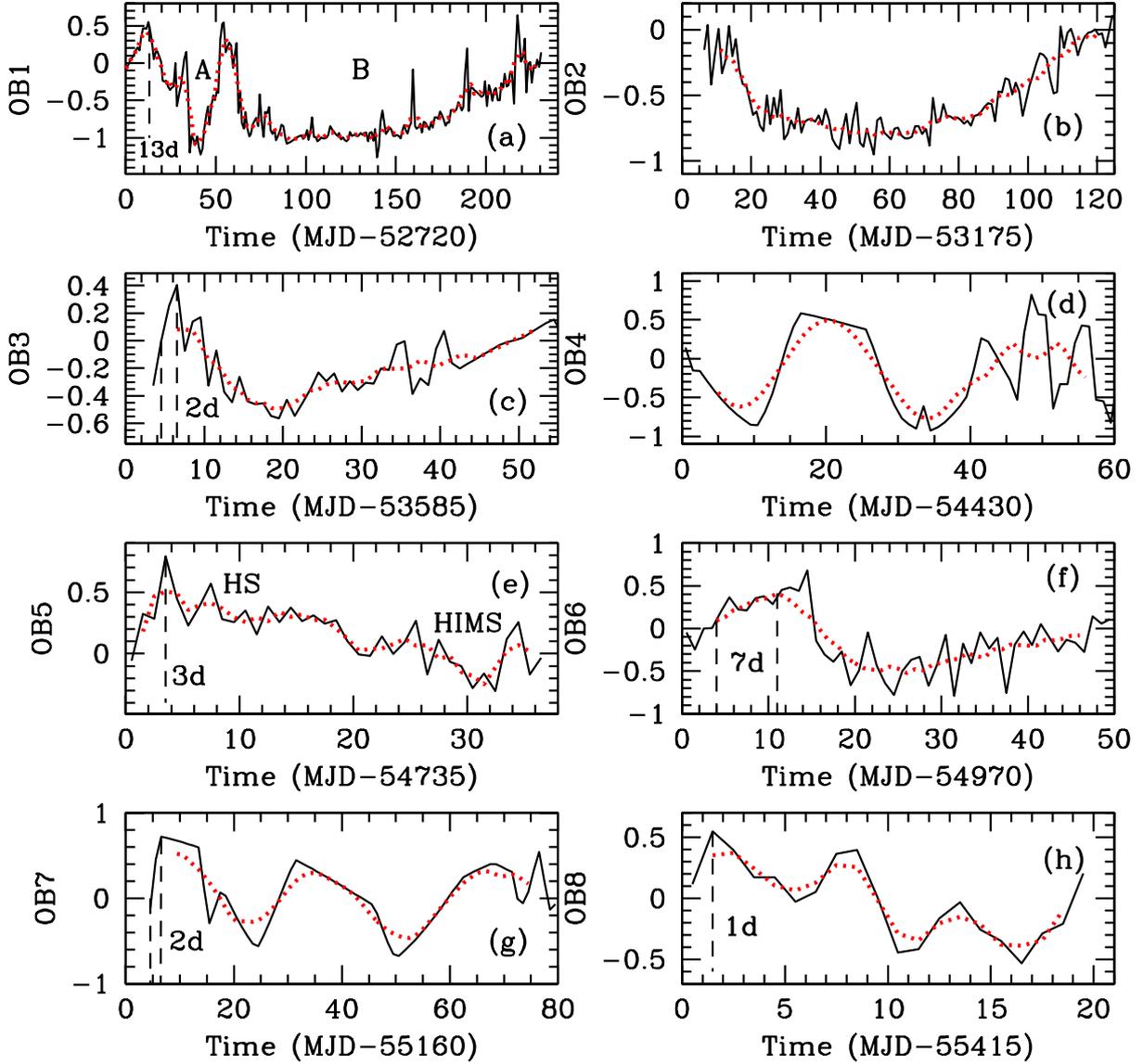}
\caption{Behaviour of $\Theta$ with time during all outbursts of Fig. 3 is shown in black/continuous. Average daily data are used. The dotted/online-red curves represent the average envelopes. With reference to the ideal diagram of $\Theta$ in Fig. 4, $\Theta$ drops down from its gradually attained highest value following the commencement of the normal outbursts OB2 (2004), OB3 (2005) \& OB6 (2009) and anomalous OB1 (2003) in (b), (c), (f) \& (a) respectively and rises again at the end; whereas its features are deviated in (d) OB4 (2007-8), (g) OB7 (2009-10) \& (h) OB8 (2010). The situation is different in (e) OB5 (2008), the trend is declining throughout; comparison with Fig. 4 readily indicates a direct HS $ \rightarrow $ HIMS spectral transition and a further hardened spectrum at the end of this {\it failed} outburst, whereas in view of (c) \& (f) it is {\it almost successful} outburst. More than one well in (a) indicates the multiplicity in OB1 or mini outbursts in OB4, OB7 \& OB8. Estimated time lags $>0d$ are marked.}
\end{center}
\end{figure}
%%%%%%%%%%%%%%%%%%%%%%%%%%%%%%%%%%%%%%%%%%%%%%%%%%%%%%%%%%
%%%%%%%%%%%%%%%%%%%%%%%%%%%%%%%%%%%%%%%%%%%%%%%%%%%%%%%%%%
\begin{figure}
\begin{center}
\includegraphics[width=\columnwidth]{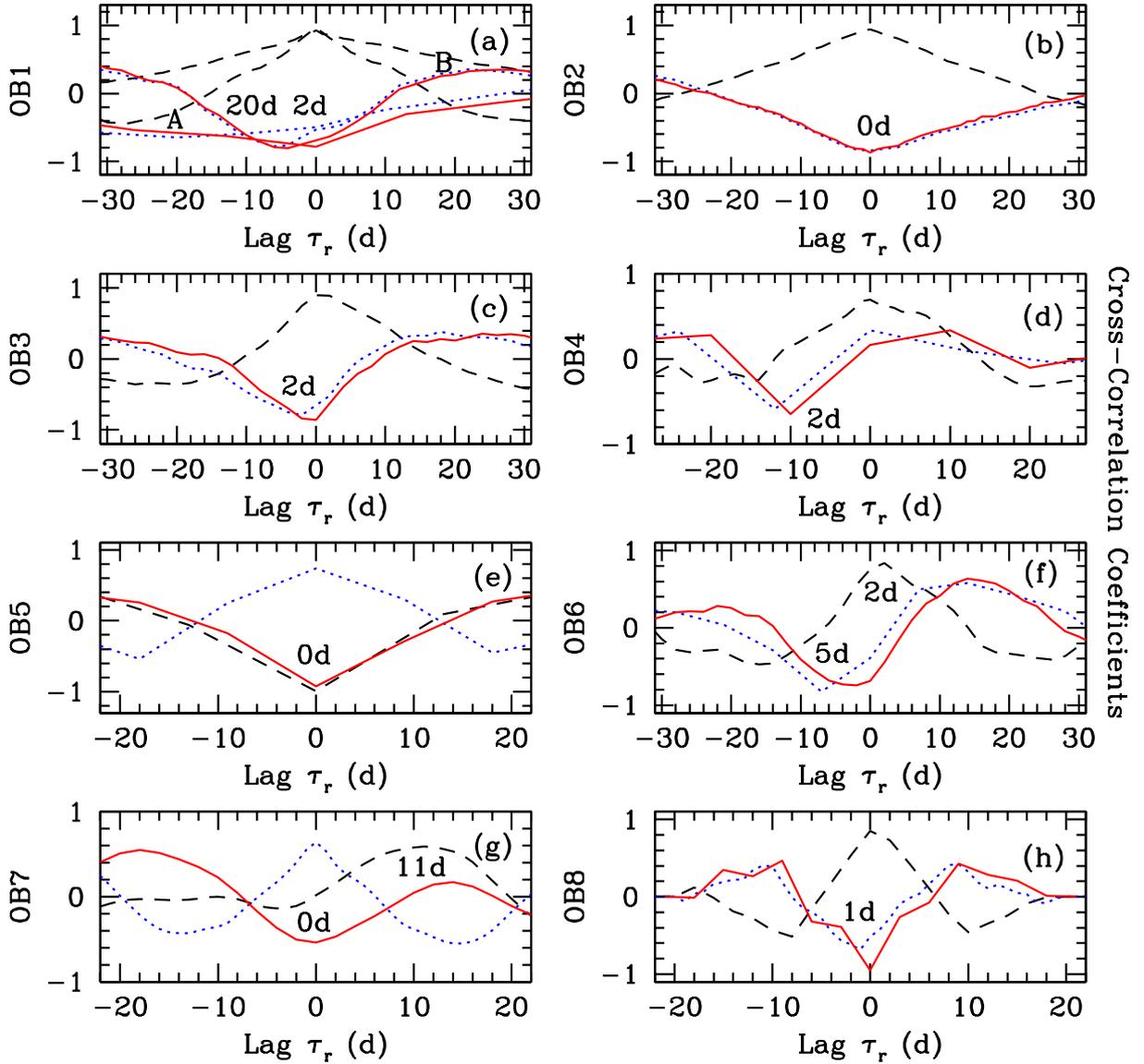}
\caption{Cross-correlation (CC) profiles of energy fluxes and $\Theta$ during outbursts are drawn. CC($\Theta$, Hard Flux) and CC($\Theta$, Soft Flux) are represented respectively by dotted/online-blue and continuous/online-red curves. Dashed black curves represent CC(Hard Flux, Soft Flux). Time lags are marked in each box. Two regions A \& B of Figs. 3a \& 5a are used separately in (a). Both (e) \& (g) are different from the rest of the plots; hard and soft fluxes are anti-correlated with reference to $\Theta$. Non-zero time lags are marked. Although the brightest outburst in (a) shows a time lag of $20(\pm8)d$, but the time lags are negligibly small in the subsequent weaker outbursts.}
\end{center}
\end{figure}
%%%%%%%%%%%%%%%%%%%%%%%%%%%%%%%%%%%%%%%%%%%%%%%%%%%%%%%%%%%%%%%%%%%%%%%%%%%%
%%%%%%%%%%%%%%%%%%%%%%%%%%%%%%%%%%%%%%%%%%%%%%%%%%%%%%%%%%%%%%%
\begin{figure}
\begin{center}
\includegraphics[width=\columnwidth]{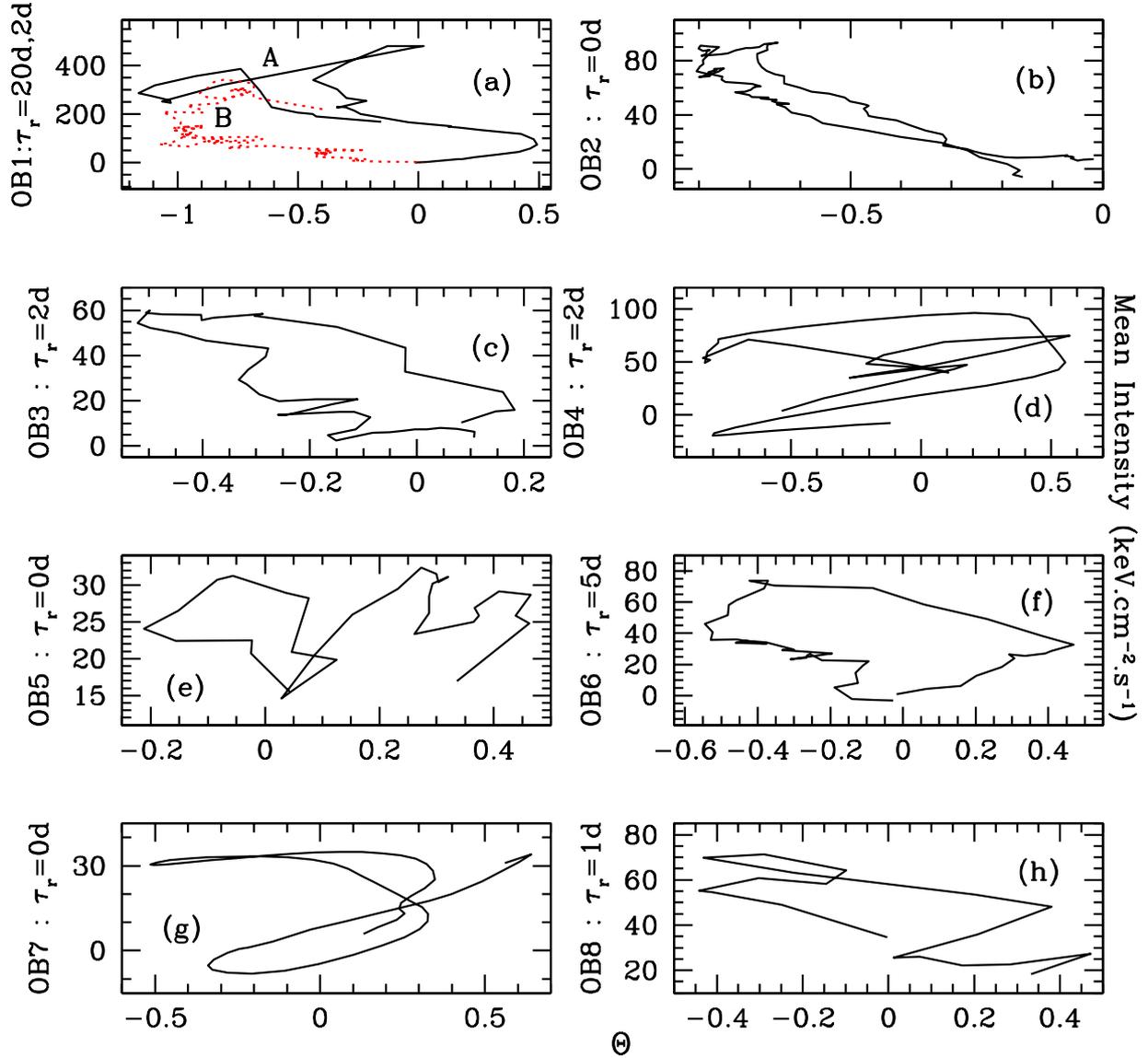}
\caption{Mean intensity in all the outbursts of Figs. 3 \& 5 is plotted with $\Theta$. Running-average data are used for clarity and comparison. Time lags computed in Fig. 6 are marked along the vertical axes. Closed ({\it hysteresis}) loops in (b), (c) \& (f) are pertaining to normal outbursts. 2-fold loop for OB1-A as well as dotted/online-red loop for OB1-B (comparable to OB2 in (b)) in (a) indicates a twin (if not a trine) outburst OB1 (2003). This is an anomalous outburst altogether. For anomalous outbursts, excursion path becomes erratic or folded (indicating multiple mini outbursts) as seen from (d), (g) \& (h). The situation is marginal for {\it almost successful / failed} outburst OB5 (2008) in (e). The biggest loop area in (a), which is consistent with the corresponding highest time lag of about $20(\pm8)d$, indicates the largest disc size in the brightest outburst OB1-A (2003) in H 1743-322. Negligibly small lags and small loop areas in all subsequent outbursts are indicative of small sizes of the Keplerian disc or nearby accumulation radii ($X_{a}$s).}
\end{center}
\end{figure}
%%%%%%%%%%%%%%%%%%%%%%%%%%%%%%%%%%%%%%%%%%%%%%%%%%%%%%%%%%%%%%%%%%%%%%%%%%%
%%%%%%%%%%%%%%%%%%%%%%%%%%%%%%%%%%%%%%%%%%%%%%%%%%%%%%%%%%%%%%%%%%%%%%%%%%%
\begin{figure}
\begin{center}
\includegraphics[width=\columnwidth]{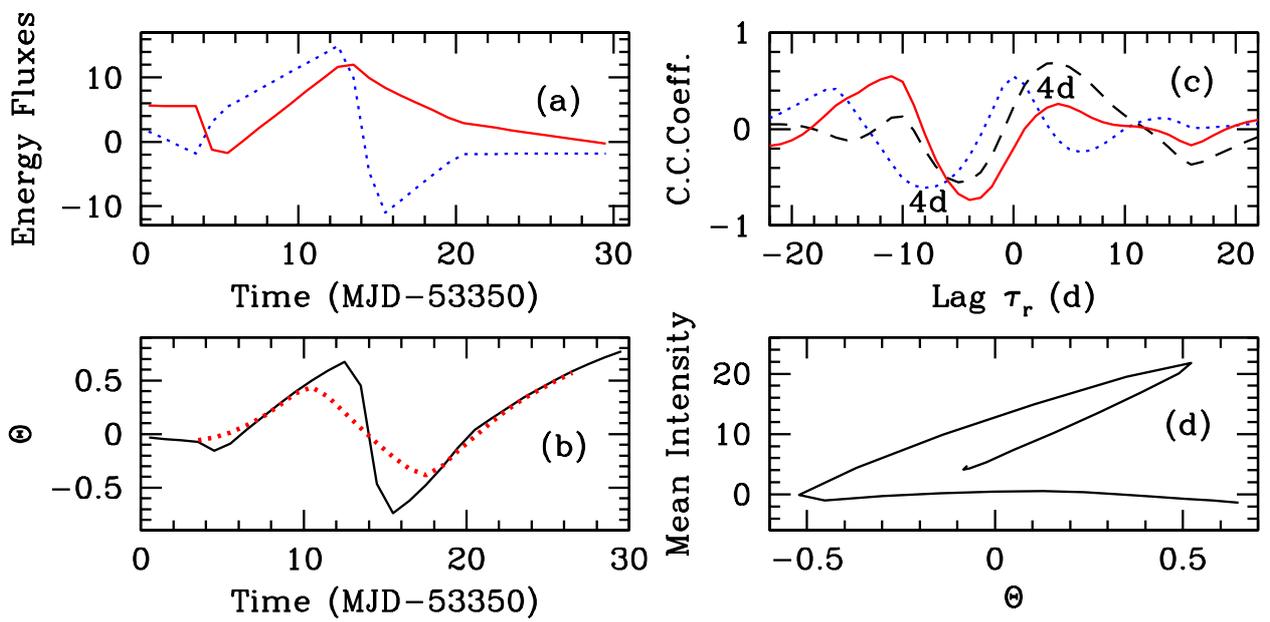}
\caption{FUSS: (a) Energy fluxes, (b) $\Theta$, (c) cross-correlation profiles, and (d) hysteresis diagram. Line types or colour stamps are the same as before. Time lag of $4d$, obtained from both flux-flux and $\Theta$-flux cross-correlations, is marked in (c).}
\end{center}
\end{figure}

%%%%%%%%%%%%%%%%%%%%%%%%%%%%%%%%%%%%%%%%%%%%%%%%%%%%%%%%%%%%%%

\end{document}